\documentclass[twocolumn,aps,prl,showpacs,amsmath,superscriptaddress]{revtex4-1}
\usepackage{amsfonts}
\usepackage{amssymb}
\usepackage{mathrsfs}
\usepackage{graphicx}
\usepackage{float}
\usepackage{xcolor}

\begin{document}

\title{Boundary-dependent Self-dualities, Winding Numbers and Asymmetrical Localization in non-Hermitian Quasicrystals}

\author{Xiaoming Cai}
\address{State Key Laboratory of Magnetic Resonance and Atomic and Molecular Physics, Wuhan Institute of Physics and Mathematics, IAPMST, Chinese Academy of Sciences, Wuhan 430071, China}

\date{\today}

\begin{abstract}

We study a non-Hermitian Aubry-Andr\'{e}-Harper model with both nonreciprocal hoppings and complex quasiperiodical potentials, which is a typical non-Hermitian quasicrystal.
We introduce boundary-dependent self-dualities in this model and obtain analytical results to describe its Asymmetrical Anderson localization and topological phase transitions.
We find that the Anderson localization is not necessarily in accordance with the topological phase transitions, which are characteristics of localization of states and topology of energy spectrum respectively.
Furthermore, in the localized phase, single-particle states are asymmetrically localized due to non-Hermitian skin effect and have energy-independent localization lengths.
We also discuss possible experimental detections of our results in electric circuits.
\end{abstract}
\maketitle

\emph{Introduction.--} Anderson localization (AL) has been a fascinating topic in condensed matter physics, ever since the classic work of Anderson in 1958 \cite{Anderson1}.
In one dimension, it is a known fact that an infinitesimal un-correlated disorder localizes all single-particle states \cite{Abrahams1}.
However, AL phase transitions can exist in one dimensional (1D) quasicrystals, such as the Aubry-Andr\'{e}-Harper (AAH) model \cite{Aubry1}, which has attracted a continuous interest both theoretically and experimentally for the past three decades \cite{Kohmoto1,Roati1,Cai1,Purkayastha1,Rossignolo1}.
The system undergoes a sudden AL phase transition at a finite strength of the quasiperiodical potential, which is guaranteed by a self-duality mapping between extended and localized phases.
%
%
Further searches for generalized self-dualities have been brought to modified AAH models with exact energy-dependent mobility edges \cite{An1,Biddle1,Ganeshan1,Wang1}.
Besides AL, the AAH model is also of the topological nature \cite{Lang1,Zhu1} and supports the Thouless pumping \cite{Kraus1}.
%

%
%
On the other hand, the ability to engineer non-Hermitian Hamiltonians, demonstrated in a series of recent experiments \cite{Zeuner1,Poli1,Peng1,Xu1,Weimann1,Pan1,Zhou1}, sparked a great interest in studying intriguing features and applications of non-Hermitian systems \cite{Feng1,Ganainy1,Miri1}.
In general, the non-Hermiticity is achieved by introducing nonreciprocal hopping or/and gain and loss, which leads to exotic phenomena, such as parity-time ($\mathcal{PT}$) phase transitions \cite{Bender1}, exceptional points \cite{Mostafazadeh1,Znojil1}, new topological invariants \cite{Gong1,Song1}, non-Hermitian skin effect and revised bulk-edge correspondence \cite{Lee1,Leykam1,Shen1,Yao1,Kunst1,Yokomizo1,Zhang2,Okuma1}.
In the presence of disorders, non-Hermitian systems can exhibit unique AL properties, such as purely imaginary disorder induced AL \cite{Freilikher1,Asatryan1,Basiri1} and non-Hermitian skin effect induced finite-strength localization-delocalization transition \cite{Hatano1,Silvestrov1,Longhi2} .
Specifically, non-Hermitian AAH models reveal remarkable impacts of the (quasi)periodical on-site potentials on the $\mathcal{PT}$ symmetry breaking \cite{Longhi3,Yuce1,Liang2,Harter1,Rivalta1}, butterfly spectrum\cite{Yuce1,Zeng2}, topological edge states \cite{Harter1,Rivalta1,Longhi4,Zeng2,Zhang3,Luo1} and localization properties of eigenstates \cite{Rivalta1,Zeng3,Longhi4,Zeng2,Longhi5,Jazaeri1,Liu1,Zeng1,Liu2}.
The interplay between nonreciprocal hopping and (quasi)periodicity gives rise to boundary-dependent topologies \cite{Zeng2,Zeng6} and localization properties \cite{Jiang1}.
Very recently,  topological nature of AL phase transition in non-Hermitian AAH models has also come to light \cite{Zeng1,Longhi1,Jiang1}.
%
%

%
So far, most studies on AL in non-Hermitian AAH models are based on numerical simulations.
Few analytical results are available due to the complex nature of energy spectrum.
In this Letter, we study topological properties and AL of non-Hermitian quasicrystals by considering a non-Hermitian AAH model with both nonreciprocal hoppings and complex quasiperiodical potentials.
We report boundary-dependent self-duality mappings, which is absent so far.
We also provide analytical results on topological phase transitions and Asymmetrical AL.
The AL phase transition is not necessarily in accordance with the topological phase transitions, which are characteristics of two aspects, localization of states and topology of energy spectrum, respectively.
Numerical calculations are also carried out to further confirm these results.

%
%
%
%
%
%
%
%

%
\emph{Non-Hermitian Aubry-Andr\'{e}-Harper model.--} We consider a 1D tight-binding non-Hermitian AAH model described by the following Hamiltonian
\begin{equation}
H=\sum_j[te^{-\eta-i\phi}a^\dagger_ja_{j+1}+te^{\eta+i\phi}a^\dagger_{j+1}a_j+V_ja^\dagger_ja_j],
\label{EQN1}
\end{equation}
where $a^\dagger_j$ ($a_j$) is the creation (annihilation) operator of a particle at site $j$.
$t_{L(R)}\equiv te^{\mp\eta}$ is the left(right)-hopping amplitude with $\eta$ characterizing the asymmetry.
$\phi$ corresponds to an applied magnetic flux or artificial gauge field.
$V_j=2V\mathrm{cos}(2\pi\beta j+\delta-ih)$ is an on-site complex quasiperiodical potential with $\beta$ an irrational number, e.g., the inverse of the golden ratio $(\sqrt{5}-1)/2$ for infinite systems.
For finite systems in numerics, $\beta=F_n/F_{n+1}$ with $F_n$ the $n$-th Fibonacci number, and the total number of lattice sites $L=F_{n+1}$.
Without loss of generality, we will take all parameters to be positive and real.
In the Hermitian limit ($\eta=h=0$) the system has a self-duality mapping and AL transition occurs at the self-duality symmetry point $V/t=1$.
Extended states for $V/t<1$ become exponentially localized when $V/t>1$ with Lyapunov exponents (LEs) $\gamma=\mathrm{ln}(V/t)$ \cite{Aubry1}, i.e., the inverse of localization lengths.
Two limit cases with $\eta=\phi=0$ and $h=\delta=0$ were considered in Ref.\cite{Longhi1,Jiang1}, respectively.
In general, non-Hermitian models may suffer from the non-Hermitian skin effect with boundary-dependent spectra and single-particle states \cite{Yao1}, and we need to study the properties under different boundaries separately.

\begin{figure}[tbp]
\begin{center}
\includegraphics[width=\linewidth, bb=74 41 571 349]{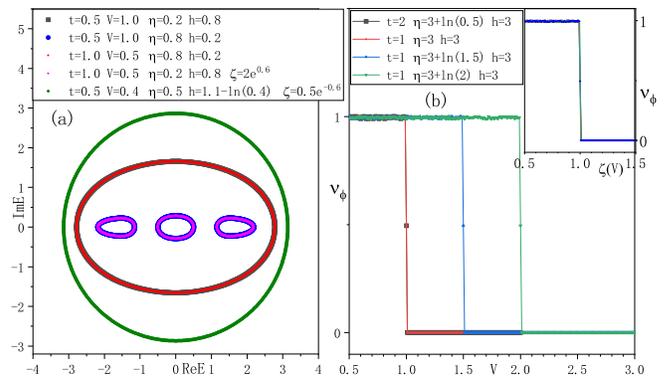}
\caption{(Color online) (a) Spectra in the complex energy plane for systems under the periodic boundary condition.
(b) Winding number $\upsilon_\phi$ vs $V$, numerically computed using Eq.(\ref{EQN2}).
Inset in (b): $\upsilon_\phi$ vs $\zeta\equiv Ve^h/te^\eta$, with the transition point $\zeta=1$.
Other parameters: $L=987$, and $\phi=\delta=0$.}
\label{Fig1}
\end{center}
\end{figure}

\emph{Self-duality, topology, and asymmetrical localization under the periodic boundary condition.--} 
Under the periodic boundary condition (PBC), we introduce a duality Fourier transformation
\begin{eqnarray}
a_j&=&\frac{1}{\sqrt{L}}\sum_{k=1}^Lb_ke^{-i k (2\pi\beta j+2\delta)-2kh},\notag\\
a^\dagger
_j&=&\frac{1}{\sqrt{L}}\sum_{k=1}^L\tilde{b}_ke^{ik (2\pi\beta j+2\delta)+2kh},
\label{EQNDuality}
\end{eqnarray}
where $b_k(\tilde{b}_k)$ is the annihilation (creation) operator in momentum space.
It takes the Hamiltonian in real space into momentum space within the same form of Eq.(\ref{EQN1}), but with simultaneous interchanges $t\leftrightarrow V$, $\eta\leftrightarrow h$ and $\phi\leftrightarrow \delta$, which defines the self-duality \cite{Sup}.
Numerical spectra under PBC support the self-duality and examples are shown in Fig.\ref{Fig1}(a).
An interesting case is obtained in the simultaneous double limit: $t\rightarrow0$, $\eta\rightarrow\infty$ with $te^{\eta}\rightarrow t'$ finite, and $V\rightarrow0$, $h\rightarrow\infty$ with $Ve^{h}\rightarrow V'$ finite.
The Hamiltonian reduces to $H'=\sum_j[t'e^{i\phi}a^\dagger_{j+1}a_j+V'e^{-i(2\pi\beta j+\delta)}a^\dagger_ja_j]$, which is a non-Hermitian half of the classic Hermitian AAH model.
The self-duality ensures the transition point $V'/t'=1$.

\begin{figure}[tbp]
\begin{center}
\includegraphics[width=\linewidth, bb=75 40 484 414]{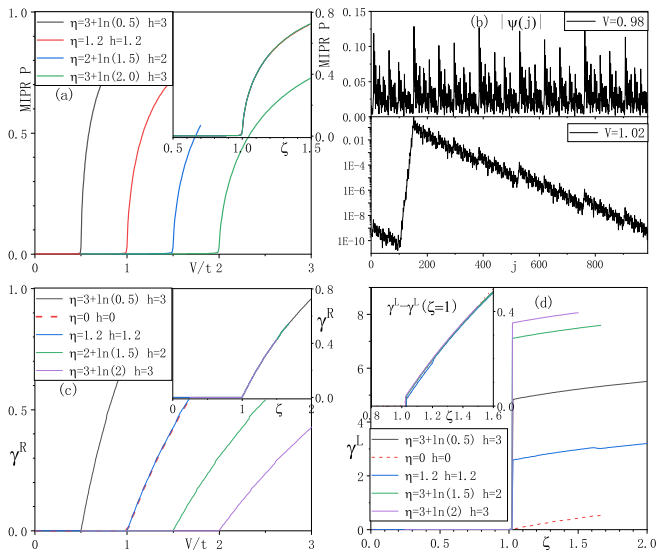}
\caption{(Color online) Anderson localization in the non-Hermitian Aubry-Andr\'{e}-Harper model under the periodic boundary condition.
(a) Mean inverse of the participation ratios (MIPRs) vs $V$.
Inset in (a): Corresponding collapsed MIPRs vs $\zeta\equiv Ve^h/te^\eta$.
(b) Two typical distributions $|\psi(j)|$ of right eigenstates for systems in extended and localized phases respectively.
$\eta=h=0.2$, in numerical calculation of (b).
(c) Mean right side Lyapunov exponents $\gamma^R$, i.e., the inverse of right side localization lengths, vs $V$.
Inset in (c): Corresponding collapsed $\gamma^R$ vs $\zeta$.
(d) Mean left side Lyapunov exponents $\gamma^L$ vs $\zeta$.
Inset in (d): $\gamma^L$ shifted by the size of jumping.
In (c), (d) and insets in them, we also show the Lyapunov exponent for the Hermitian Aubry-Andr\'{e}-Harper model (red dash lines).
Other parameters: $L=987$, $t=1$ and $\phi=\delta=0$.}
\label{Fig2}
\end{center}
\end{figure}

Due to connection to the two dimensional quantum Hall system, topological properties of 1D superlattices and quasicrystals in Hermitian \cite{Lang1,Zhu1} and non-Hermitian \cite{Jiang1,Longhi1,Zeng1} systems have attracted great interest recently, where some parameter (such as $\delta$ or $\phi$ in our case) is treated as an additional dimension.
In the same spirit, here we adopt the winding numbers of energy spectra \cite{Longhi1,Jiang1,Zeng1}
\begin{equation}
\upsilon_{\tau}=\lim\limits_{L\rightarrow\infty}\frac{1}{2\pi i}\int_0^{2\pi/L}d\tau\frac{1}{\partial\tau}\mathrm{log}[\mathrm{det}(H)],
\label{EQN2}
\end{equation}
with $\tau=\phi$ or $\delta$, which refer to the widely used winding numbers evaluated by applying a magnetic flux $\phi$ and the phase $\delta$ in the on-site potential respectively.
Analytically derived in the Supplementary Material \cite{Sup}, under the PBC, two winding numbers of energy spectra are related and
\begin{equation}
\upsilon_\phi=1-\upsilon_\delta=\theta(\frac{te^\eta}{Ve^h}-1),
\label{EQN3}
\end{equation}
with $\theta(x)$ the step function.
The topological phase transition is also verified by numerical calculations of winding numbers directly using Eq.(\ref{EQN2}).
Examples of numerical $\upsilon_\phi$ vs $V$ are presented in Fig.\ref{Fig1}(b).
After rescaling, all curves collapse with the precise topological phase transition point $\zeta\equiv Ve^h/te^\eta=1$.
Eq.(\ref{EQN3}) suggests a more general self-duality $\zeta\leftrightarrow 1/\zeta$, which is not supported by numerical results [see Fig.\ref{Fig1}(a)].
With an irrational $\beta$, the quasiperiodical potential acts as a quasirandom disorder which induces the localization of single-particle eigenstates.
We denote the right eigenstates of $H$ by $|\Psi^R_s\rangle=\sum_j\psi_s(j)a^\dagger_j|0\rangle$ with $s$ the index of eigenstates.
The localization can be characterized by the inverse of the participation ratio (IPR) $P_s=\sum_j|\psi_s(j)|^4/[\sum_j|\psi_s(j)|^2]^2$.
For a localized state the IPR approaches to around $1$, whereas for an extended state the IPR is of the order $1/L$.
In order to characterize the localization of the system we define the mean inverse of the participation ratio (MIPR) $P=\sum_sP_s/L$.
We show MIPRs vs $V$ for different systems under the PBC in Fig.\ref{Fig2}(a) and corresponding rescaled ones in the inset.
All curves collapse with the AL phase transition point $\zeta=1$, which is the same as the topological phase transition point.
No mobility edge is encountered.
All eigenstates are extended when $\zeta<1$, whereas localized when $\zeta>1$. %
One can also define (M)IPR for the left eigenstates, which give the same conclusion.

\begin{figure}[tbp]
\begin{center}
\includegraphics[width=\linewidth, bb=71 48 443 412]{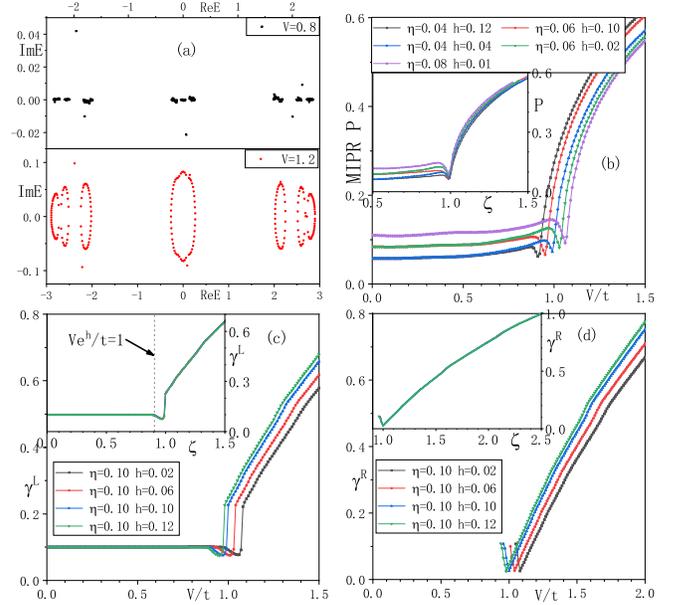}
\caption{(Color online) Spectra and Anderson localization for the non-Hermitian Aubry-Andr\'{e}-Harper model under the open boundary condition.
(a) Two typical spectra in complex energy plane with different winding numbers $\upsilon_\delta$.
$\eta=h=0.1$, in numerical calculation of (a).
(b) Mean inverse of the participation ratios (MIPRs) vs $V$.
Inset in (b): Corresponding MIPRs vs $\zeta$.
(c,d) Mean left/right side Lyapunov exponents $\gamma^{L(R)}$ vs $V$.
Insets in (c,d): Corresponding collapsed $\gamma^{L(R)}$ vs $\zeta$.
Numerical results before transition points in (d) and inset are not reliable, because of the skin effect induced right-boundary-localization nature of right eigenstates.
Other parameters: $L=233$, $t=1$ and $\phi=\delta=0$.}
\label{Fig3}
\end{center}
\end{figure}

In order to explore details of the localization, in Fig.\ref{Fig2}(b) we show two typical distributions of right eigenstates in extended and localized phases, respectively.
Distinctively, under the PBC the localized state has an asymmetrical exponential decay.
We adopt the asymmetrical wave functions
\begin{equation}
\psi_s(j)\propto\left\{\begin{array}{c}e^{-\gamma^R_s(j-j_0)},j>j_0,\\e^{-\gamma^L_s(j_0-j)},j<j_0,\end{array}\right.
\label{EQN4}
\end{equation}
which manifest different exponential decaying behaviours on both sides of the localization center $j_0$ with two LEs $\gamma^{R(L)}_s$.
Extracted by fitting the numerical data with Eq.(\ref{EQN4}), the mean right and left side LEs $\gamma^{R(L)}=\sum_s\gamma^{R(L)}_s/L$ are presented in Fig.\ref{Fig2}(c)(d), along with the LE for classic Hermitian AAH model.
All mean right side LEs $\gamma^{R}$ collapse into a single curve with the AL phase transition point $\zeta=1$.
Based on the known LE $\gamma=\mathrm{ln}(V/t)$ for the Hermitian AAH model, the right side LE $\gamma^{R}=\mathrm{ln}(Ve^h/te^\eta)$ for the non-Hermitian AAH model under the PBC.
On the other side, the mean left side LE experiences a sudden jump at $\zeta=1$.
The size of jumping only depends on and approximately equals $2\eta$.
After shifting the jump, all mean left side LEs collapse into the LE for Hermitian AAH model, which are shown in the inset of Fig.\ref{Fig2}(d).
Both left and right side LEs are energy-independent.
In a word, under the PBC right eigenstates in the localized phase have LEs $\gamma^R=\mathrm{ln}(Ve^h/te^\eta)$,
$\gamma^L=\gamma_0+\gamma^R$, and $\gamma_0\simeq 2\eta$.
\emph{Skin effect induced novel physics under the open boundary condition.--} Under the open boundary condition (OBC), the model with a non-zero $\eta$  suffers from the non-Hermitian skin effect \cite{Gong1}.
In the absence of disorder, all right eigenstates are exponentially localized at the right(left) end when $\eta>(<)0$.
In general, we introduce an asymmetric similarity transformation $a_j=e^{(\eta+i\phi)j}b_j$ and $a^\dagger_j=e^{-(\eta+i\phi)j}\tilde{b}_j$.
The Hamiltonian Eq.(\ref{EQN1}) is mapped to
\begin{equation}
H_1=\sum_j[t\tilde{b}_jb_{j+1}+t\tilde{b}_{j+1}b_j+V_j\tilde{b}_jb_j].
\end{equation}
Then Hamiltonians $H$ and $H_1$ have the exact same spectra and winding numbers $\upsilon_{\phi(\delta)}$ under the OBC.
Note that here we focus on bulk properties of the system, and these two winding numbers are not useful to predict topological edge states.
Please refer to the Supplementary Material for a brief study of edge states \cite{Sup}.
For the moment, we concentrate on the Hamiltonian $H_1$, which was originally introduced in Ref.\cite{Longhi1} (see also in the Supplementary Material \cite{Sup}).
The model supports a topological AL phase transition at $Ve^h/t=1$.
The winding numbers \cite{Sup}
\begin{equation}
\upsilon_\phi=0,\upsilon_\delta=\theta(Ve^h/t-1),
\end{equation}
because $H_1$ does not depend on $\phi$.
This model, with $\eta=0$, does not suffer from the skin effect.
All bulk states are extended when $Ve^h/t<1$.
When $Ve^h/t>1$ all bulk states are exponentially localized with energy-independent and equal right and left side LEs $\gamma_1=\mathrm{ln}(Ve^h/t)$ \cite{Sup}.
Back to the Hamiltonian $H$, under the OBC the model does not have a self-duality when $h\neq0$.
When $Ve^h/t<1$, $\upsilon_\delta=0$, and the spectrum mainly consists of three "bands" without loop.
While when $Ve^h/t>1$, $\upsilon_\delta=1$, and the spectrum consists of loops where the complex spectral trajectory encircles the origin [see Fig.\ref{Fig3}(a) for examples].
There can not be any self-duality transformation mapping between spectra with different topologies.
However, when $h=0$, based on the relation to the Hermitian AAH model, one can find a self-duality transformation \cite{Sup}
\begin{eqnarray}
a_j&=&\frac{1}{\sqrt{L}}e^{i(\phi-\delta)j+\eta j}\sum_kb_ke^{-ik(2\pi\beta j+\delta+\phi)-k\eta},\notag\\
a^\dagger
_j&=&\frac{1}{\sqrt{L}}e^{-i(\phi-\delta)j-\eta j}\sum_k\tilde{b}_ke^{ik(2\pi\beta j+\delta+\phi)+k\eta}.
\label{EQNDuality1}
\end{eqnarray}
It interchanges $t$ and $V$ in Hamiltonian $H(h=0)$.
The spectrum of $H(h=0)$ is the same as for the Hermitian AAH model, with the self-duality critical point $V/t=1$, which is not the localization critical point \cite{Jiang1}.

On the other hand, following the asymmetric similarity transformation between $H$ and $H_1$, under the OBC a right eigenstate of $H$ satisfies $\psi_s(j)=e^{(\eta+i\phi)j}\psi'_s(j)$ with $\psi'_s(j)$ the corresponding right eigenstate of $H_1$.
It clearly shows how the skin effect affects states in different phases: For extended states of $H_1$, the corresponding wave functions $\psi_s(j)$ are localized at the right end with left side LEs $\gamma^L=\eta$; For localized states of $H_1$, wave functions $\psi_s(j)$ have the form
\begin{equation}
\psi_s(j)\propto\left\{\begin{array}{c}e^{-(\gamma_1-\eta)(j-j_0)},j>j_0,\\e^{-(\gamma_1+\eta)(j_0-j)},j<j_0.\end{array}\right.
\end{equation}
Then the condition $\gamma_1-\eta>0$ indicates the localization of right eigenstates, and $\gamma_1-\eta=0$ gives the AL phase transition point
\begin{equation}
\zeta\equiv Ve^h/te^\eta=1.
\end{equation}
It is the same as the transition point under the PBC, but not the transition point for spectrum or topology under the OBC.
Furthermore, in the localized phase, the right and left side LEs
\begin{eqnarray}
\gamma^R&=&\gamma_1-\eta=\mathrm{ln}(Ve^h/te^\eta),\notag\\
\gamma^L&=&\gamma_1+\eta=2\eta+\gamma^R.
\end{eqnarray}
The ALs under both boundary conditions are identical, proving the insensitivity of the localized states to boundaries.

The above analytical results are in good agreement with numerical ones.
In Fig.\ref{Fig3}(b) we show MIPRs under the OBC.
Due to the skin effect induced boundary-localization nature of right eigenstates, the AL phase transition point should correspond to the most extended case with the smallest MIPR.
Therefore, there are deep dives in MIPRs around $\zeta=1$.
In Fig.\ref{Fig3}(c) we show mean left side LEs which show a jump to $2\eta$ at the transition point $\zeta=1$.
When $\zeta<1$, $\gamma^L=h$ owing to the skin effect, except there is a dive just before the transition point which also indicates the delocalization of states.
The dive begins at $Ve^h/t=1$, which is the topological phase transition point or the AL phase transition point for the model $H_1$.
This skin effect induced phase was called skin phase \cite{Jiang1}.
In Fig.\ref{Fig3}(d) we show mean right side LEs, which show a clear transition at $\zeta=1$ and correct LEs in localized phase in spite of the unreliable numerical LEs in skin phase due to the right-boundary-localization nature of right eigenstates.

\emph{Experimental realization.--} The physics of non-Hermitian AAH model [Eq.(\ref{EQN1})] can be simulated by electric circuits \cite{Zeng1}, which recently have turned out to be powerful platforms to simulate non-Hermitian and/or topological phases \cite{Ashida1}.
The single-particle eigenvalue problem is simulated by the Kirchhoff's current law $I_a=\sum_{b=1}^LJ_{ab}V_b$, where $J$ the Laplacian of circuit acts as the effective Hamiltonian, and $I_a$ and $V_a$ are the current and voltage at node $a$.
On-site complex potentials are provided by grounding nodes with proper resistors \cite{Schindler1}, and asymmetrical hopping amplitudes are realized by negative impedance converters with current inversion (INICs) \cite{Hofmann1}.
Furthermore, the boundary-dependent spectra could be obtained by measuring two-node impedances \cite{Zeng2}.
%

%
%
%
%
%
%
%
In summary, we have analytically studied the non-Hermitian AAH model with both nonreciprocal hoppings and complex quasiperiodical potentials.
We first report boundary-dependent self-dualities between localized and extended phases.
We also provide analytical results on topological phase transitions and asymmetrical AL.
The AL phase transition is not necessarily in accordance with the topological phase transitions, which are characteristics of two aspects, localization of states and topology of energy spectrum, respectively.
Under weak disorders, the skin effect dominates and the system exhibits boundary-dependent behaviours.
In the localized phase, states are asymmetrically localized with two LEs.
Analytical forms of the energy-independent LEs are derived.
We also demonstrate that in non-Hermitian systems the AL is insensitive to boundary conditions.
Physics shown above can be experimentally studied in electronic circuits.
In the future, it would be interesting to extend the present study to other non-Hermitian quasicrystals, such as the Fibonacci lattice, ladders, and systems with mobility edges.
This work is supported by the NKRDP under Grant No. 2016YFA0301503, the key NSFC grant No.\ 11534014 No.\ 11874393 and No.\ 1167420, and the National Key R\&D Program of China  No.\ 2017YFA0304500.

\end{document}